\documentclass[natbib, final]{svjour3}    
\usepackage[utf8]{inputenc}
\usepackage{a4wide}
\usepackage{amsmath, bm}
\usepackage{amssymb}
\usepackage{amsfonts}
\usepackage{graphicx}
\usepackage{epstopdf}
\usepackage{booktabs}
\usepackage{dutchcal}
\usepackage{color, colortbl}
\usepackage{caption}
\usepackage[]{hyperref}

\journalname{Celestial Mechanics and Dynamical Astronomy}
 \title{Long-term Attitude Dynamics of Space Debris in Sun-synchronous Orbits: Cassini Cycles and Chaotic Stabilization}
 \titlerunning{Long-term Attitude Dynamics of Space Debris}

 \author{S. Efimov \and
         D. Pritykin \and
         V. Sidorenko}
 \institute{ S. Efimov
           \at Moscow Institute of Physics and Technology \\
					 \smallskip
                9 Institutskiy per., Dolgoprudny, Moscow Region, 141701, Russian Federation	\\
					 \smallskip
                \email{efimov.ss@phystech.edu}
           \and D. Pritykin
           \at Moscow Institute of Physics and Technology \\
						\smallskip
                9 Institutskiy per., Dolgoprudny, Moscow Region, 141701, Russian Federation	\\
           \smallskip
           \and V. Sidorenko
           \at Keldysh Institute of Applied Mathematics\\
					     Russian Academy of Sciences, \\
					 \smallskip
               Miusskaya Sq., 4, 125047 Moscow, Russian Federation \\
						and \\	
					 \at Moscow Institute of Physics and Technology	\\
					 \smallskip
                9 Institutskiy per., Dolgoprudny, Moscow Region, 141701, Russian Federation}											
 \authorrunning{S.Efimov et al.}
 \date{}				

\begin{document}
\bibliographystyle{plainnat}
 \maketitle
 \begin{abstract}

Comprehensive analysis of space debris rotational dynamics is vital for active debris removal missions that require physical capture or de-tumbling of a target. We study the attitude motion of used rocket bodies acknowledgedly belonging to one of the categories of large space debris objects that pose an immediate danger to space operations in low Earth orbits. Particularly, we focus on Sun-synchronous orbits (SSO) with altitudes in the interval $600\div800$~km, where the density of space debris is maximal. Our mathematical model takes into account the gravity gradient torque and the torque due to eddy currents induced by the interaction of conductive materials with the geomagnetic field. Using perturbation techniques and numerical methods we examine the deceleration of the initial fast rotation and the subsequent transition to a relative equilibrium with respect to the local vertical. A better understanding of the latter phase is achieved owing to a more accurate model of the eddy currents torque than in most prior research. We show that SSO precession is also an important factor influencing the motion properties. One of its effects is manifested at the deceleration stage as the angular momentum vector oscillates about the direction to the south celestial pole.


\end{abstract}

\section{Introduction}\label{intro}

This paper presents a study of rotational dynamics of large space debris objects in Sun-synchronous orbits. SSO are characterized by 600--800~km altitude and inclination of about $90^{\circ}$ (\cite{V2007}). These orbits are best suited for the Earth' observation from space, because of consistent lighting conditions in their subsatellite points for all satellite passes. Throughout the last few decades SSO have been in use, there amassed quite a number of large debris objects, posing a real threat to space activities. At present, the SSO region is characterized by the highest debris density and requires to be cleaned (\cite{anselmo2016}). Different aspects of active debris removal (ADR) missions are brought up in \cite{B2013}, \cite{V2014}. One of the generally accepted ADR scenarios is tugging debris objects to the lower orbits, whereupon they burn in the atmosphere or fall to the Earth (\cite{A2013}). Most ADR techniques depend substantially on the character of the debris object's rotational dynamics, hence much effort has been spent lately to determine the rotation parameters through ground-based observations (\cite{K2016, K2014, L2013, silha2017, santoni2013, yanagisawa2012}). At the same time, much attention has been paid to studying space debris rotational dynamics theoretically (\cite{G2015, L2015, O2012, P2012, albuja2015, sagnieres2017}).

According to observation data (\cite{silha2017}), there are two major types of large debris objects -- defunct satellites and rocket bodies. Although much of what is discussed in this paper regarding the long-term attitude motion evolution is applicable to both classes of debris objects, there are also distinctions, which require separate treatment. For this reason we shall here confine ourselves to the dynamics of the rocket bodies, whereas the defunct satellites story is told in (\cite{defsat2017}).

Simulation of rotational dynamics for a typical object of the rocket bodies class (Ariane 4 H10 stage) is conducted in (\cite{P2012}). The model we use in our study comprises the same key factors as in (\cite{P2012, G2015}) -- gravity gradient torque and the torque due to eddy currents. As did \cite{L2015} we also take into account the orbit precession, which is responsible for remarkable dynamical effects unexamined in previous studies. Besides that, when calculating the torque due to eddy currents we employ a more accurate formula for eddy currents torque proposed in \cite{G1972} and \cite{M1985}, which includes terms describing the influence of orbital motion that are considered small for fast rotations and are often neglected. The fact that rotational dynamics at $500-1000$~km altitudes is substantially influenced by torques due to eddy currents became clear immediately after the first artificial Earth satellites launches (e.g., \cite{S1964, O1967}). It so happened, however, that when dealing with this phenomenon many researchers were mainly interested in fast rotations, whose orbital period is significantly greater than the rotation period. The complete formula for eddy currents torque allows correct description of all stages of the rotational dynamics evolution, including that when the angular velocity is comparable to the mean motion. Moreover, even for relatively large angular velocities (10--50 times greater than the mean motion) these terms can cause significant changes in the rotational axis direction for prograde spins. As in prior research we neglect other environmental torques, which can be done for the chosen class of objects.

It turns out that the attitude motion  evolution can be divided into three stages: transition to the rotation about the axis with the greatest moment of inertia (so called ``flat'' or ``principal axis'' rotation), exponential deceleration of angular velocity, and the stage of temporary slow chaotic dynamics. During the first relatively brief stage, the motion is primarily determined by internal dissipation. In the second stage, angular velocity decays exponentially due to eddy currents. When the angular velocity becomes comparable to the mean motion, the attitude dynamics begins to seem chaotic. This chaos, however, is temporary in the case of rocket bodies dynamics. It results typically in the stable relative equilibrium of the object with respect to local vertical (more exactly, the final regime corresponds to small oscillations about relative equilibrium).

The paper is organized as follows. Section~\ref{sec:model} describes the main assumptions of our model and the equations for gravity gradient torque and torque due to eddy currents. Section~\ref{sec:analytic} presents the analytical study of the debris objects attitude motion evolution. We derive the evolution equations and introduce the means of their geometric interpretation in terms of angular momentum direction. At the end of this section we also provide the classification of the long-term evolution scenarios. Section~\ref{sec:numeric} contains the simulation results validating the conclusions drawn from the analytical study and providing an understanding of the system's characteristic behavior in the stage of temporary chaos. Finally, the last section summarizes the results obtained for the characteristic evolution of large debris object rotational dynamics in SSO.

\section{Mathematical model of a debris object rotational dynamics in SSO}\label{sec:model}
Consider an object in a circular geocentric orbit of radius $R_O$ and inclination $i$. The Earth's oblateness causes the orbit's precession with angular velocity
\begin{equation*}
n_{\Omega}\approx-\frac{3 J_2 \mu_{G}^{1/2}R_{E}^2}{2 R_O^{7/2}}\cos{i},
\end{equation*}
where $R_E=6378.245$~km is the Earth's mean equatorial radius, $\mu_G=3.986\cdot10^5$~km$^3$/s$^2$ is the gravity parameter of the Earth, $J_2 =1.082626\cdot10^{-3}$ is the first zonal harmonic coefficient in the expansion of the Earth's gravity field. Our model pertains to SSO, where $\cos{i}<0$ and, consequently, $n_\Omega>0$, i.e. the longitude of ascending node increases.

Argument of latitude $u$ varies as a linear function of time:
\begin{equation*}
\dot{u}=\omega_D,
\end{equation*}
where $\omega_D = 2\pi/T_D$, $T_D$ is the draconic period of the object's revolution around the Earth (the time between two consecutive passages through the ascending node). Employing the formula for draconic period, given in \cite{V2007}, we obtain:
\begin{equation}
\omega_D=\omega_o\left[ 1 - \frac{3}{2}{J_2}{{\left( {\frac{{{R_E}}}{R_O}} \right)}^2}(1 - 4{{\cos }^2}i) \right],
\label{eq:omegaD}
\end{equation}
where $\omega_o$ is the mean motion for the circular orbit of radius $R_O$ in the central gravity field with parameter $\mu_G$.

Let us assume that the ellipsoid of inertia of the considered object is close to elongated ellipsoid of rotation. This assumption holds for rocket bodies, which are the primary target of this study. As in prior research (\cite{G2015, L2015, P2012}) when modeling the rotational dynamics with respect to object's center of mass, we shall take into account gravity gradient torque $\mathbf{M}_{G}$ and torque due to eddy currents $\mathbf{M}_{EC}$.

Gravity gradient torque acting on the object in the Earth's gravity field is given by the formula (\cite{B1966}):
\begin{equation*}
\mathbf{M}_{G} = \frac{3\mu_G}{R_O^5} \mathbf{R}_O \times \mathbf{J} \mathbf{R}_O,
\end{equation*}
where $\bf{J}$ is the inertia tensor of the object, $\mathbf{R}_O$ is the vector from the center of the Earth to the object's center of mass $O$.

Torque due to eddy currents can be expressed as (\cite{G1972, M1985}):
\begin{equation}
\mathbf{M}_{EC} = -\mathbf{B} \times \mathbf{S}(\bm{\omega} \times \mathbf{B} - \dot{\mathbf{B}}),
\label{eq:MEC}
\end{equation}
where $\mathbf{S}$  is the magnetic tensor of the object, $\mathbf{B}$ is the magnetic field, the derivative $\dot{\mathbf{B}}$ is calculated in a non-rotating reference frame with the origin at point $O$.

Geomagnetic field is modeled as a field of dipole placed into the center of the Earth:
\begin{equation*}
\mathbf{B} = \frac{\mu_0\mu_E}{4\pi R_O^3}\left [ \frac{3\mathbf{R}_O(\mathbf{k}_E,\mathbf{R}_O)}{R_O^2}-\mathbf{k}_E\right],
\end{equation*}
where $\mu_0 \approx 1.257\cdot10^{-6}$~N$\cdot$A$^{-2}$ is the magnetic constant, $\mu_E \approx 7.94\cdot10^{22}$~A$\cdot$m$^2$ is the Earth's magnetic dipole moment, $\mathbf{k}_E$ is the dipole direction.

In Section~\ref{sec:analytic}, where the evolution equations are derived, we assume for simplicity that the dipole is directed along the Earth's rotation axis (``axial'' dipole model). In Section~\ref{sec:numeric} we validate this assumption by carrying out simulations with the use of a more precise model (``inclined'' dipole, making an angle $\delta_\mu=11^{\circ}33'$ with the Earth's rotation axis). It is shown in the Section~\ref{sec:validation} that within the accuracy of the averaging procedure the dipole model simplification is valid and allows studying the secular effects in the object's motion using the evolution equations obtained for the ``axial'' dipole model.

The initial motion is assumed to be a rotation about the axis with the greatest moment of inertia. The initial angular velocity absolute value is specified in Section~\ref{subsection:map} and assumed to be much greater than the mean motion $\omega_o$. Such regime sets in quite fast under the influence of internal dissipation due to the motion of residual fuel in the fuel tanks of the rocket body (\cite{O2012, dycoss2017}).

The parameters used in simulations are listed in Table~\ref{tbl:params}.

\begin{table}[hb]
\begin{center}
\captionsetup{justification=centering}
\caption{Rocket bodies' parameters}\label{tbl:params}
\begin{tabular}{lll}
\toprule
Inertia tensor components    & \hspace{0.5cm}$A$                 & \hspace{1cm}10815 kg$\cdot$m$^2$  \\
														 & \hspace{0.5cm}$B$                 & \hspace{1cm}10739 kg$\cdot$m$^2$  \\
														 & \hspace{0.5cm}$C$                 & \hspace{1cm}1441 kg$\cdot$m$^2$   \vspace{0.1cm}\\
Magnetic tensor components   & \hspace{0.5cm}$S_{x'x'}=S_{y'y'}$ & \hspace{1cm}$2.18\cdot10^6$ S$\cdot$m$^4$\\
														 & \hspace{0.5cm}$S_{z'z'}$          & \hspace{1cm}$1.32\cdot10^6$ S$\cdot$m$^4$  \vspace{0.1cm}\\
Orbit altitude    & \hspace{0.5cm}$H$                 & \hspace{1cm}770 km \vspace{0.1cm}\\
Orbit inclination & \hspace{0.5cm}$i$                 & \hspace{1cm}98.7$^\circ$ \\
\bottomrule
\end{tabular}
\end{center}
\end{table}

\section{Analytical study of fast rotations evolution}\label{sec:analytic}

\subsection{Reference frames}

We shall use several reference frames with the common origin in the object's center of mass $O$.

\begin{figure}[htb]

\center{\includegraphics[width=80mm]{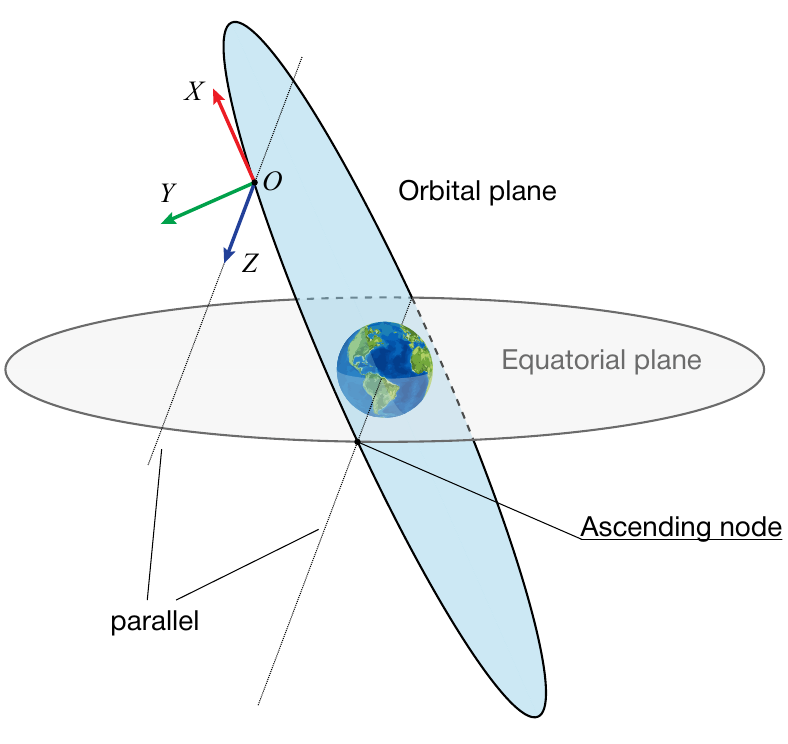}}

\caption{Semi-orbital reference frame}
\label{fig:reframeXYZ}
\end{figure}

$OXYZ$ is a semi-orbital reference frame: axis $OY$ is perpendicular to the orbital plane, axis $OZ$ is parallel to the vector from the Earth's center to the ascending node, axis $OX$ is directed along the object's center of mass velocity as it passes the ascending node (Fig.~\ref{fig:reframeXYZ}).

\begin{figure}[htb]

\center{\includegraphics[width=100mm]{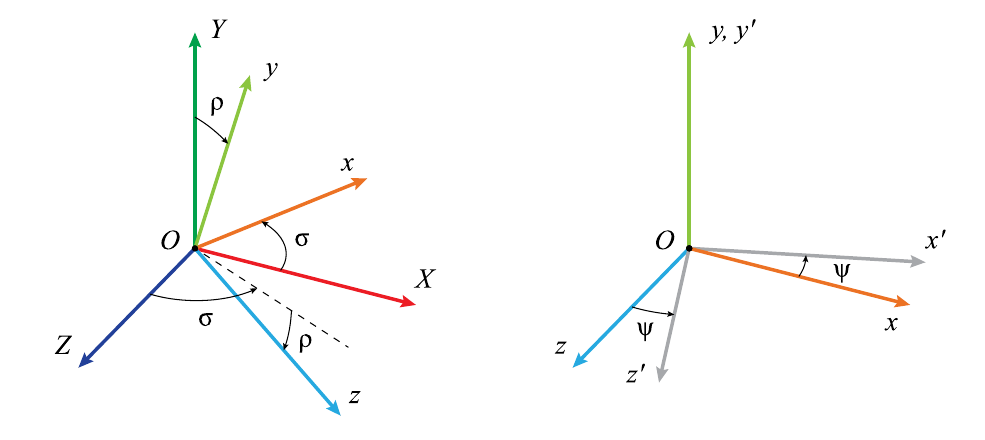}}

\caption{Orientation of reference frame $Oxyz$ relative to $OXYZ$ (left) and orientation of reference frame $Ox'y'z'$ relative to $Oxyz$ (left)}
\label{fig:reframesxyz}
\end{figure}

$Oxyz$ is a reference frame bound to the vector of the object's angular momentum with respect to its center of mass $\mathbf{L}$: axis $Oy$ goes along $\mathbf{L}$, axis $Ox$ lies in the orbital plane (Fig.~\ref{fig:reframesxyz}). The attitude of $Oxyz$ with respect to $OXYZ$ is described by angles $\rho$, $\sigma$ (let us note, that given the values of these angles, we define the direction of the angular momentum $\mathbf{L}$ as well).

$Ox'y'z'$ is a body-fixed frame with  the axes directed along the object's principal axes of inertia. For simplicity we neglect in this section the small asymmetry of the object. Thus, the inertia tensor with respect to $Ox'y'z'$ is diagonal:
\begin{equation}\label{diagonalInertia}
\mathbf{J}=diag(A,A,C), \quad A>C.
\end{equation}

\emph{Remark}: The assumption of dynamical symmetry is not restrictive. Secular evolution of the attitude motion in the case of triaxial ellipsoid of inertia is described by exactly the same equations with slightly modified parameters (see Section~\ref{subsec:avcas}).

Numeric experiments show that during the stage of exponential deceleration vector $\mathbf{L}$ remains virtually perpendicular to the object's symmetry axis. It helps simplifying the mathematical model: we further assume that $Oy'$ is always directed along $\mathbf{L}$ and thus coincides with $Oy$ (this approach allows rigorous justification, which is omitted here). Let $\psi$ be a rotation angle around $Oy$, which describes the attitude of the body frame $Ox'y'z'$ with respect to $Oxyz$. When $\psi=0$ the two frames coincide with each other~(Fig.~\ref{fig:reframesxyz}).

Let us denote the unit vectors of the introduced reference frames by $\mathbf{e}_{\xi}$, where the lower index $\xi$ refers to the corresponding coordinate axis $\xi \in \{X,...,x...,x',...\}$. The unit vector $\mathbf{e}_{y}$ can also be denoted by $\mathbf{e}_{L}$ to emphasize that it is directed along $\mathbf{L}$.

Let us introduce two transformation matrices:
\begin{equation*}
\mathbf{\Gamma}'=
\begin{bmatrix}
    \cos{\psi}   & 0 & -\sin{\psi} \\
    0            & 1 & 0 \\
    \sin{\psi}   & 0 & \cos{\psi}
\end{bmatrix}, \quad
\mathbf{\Gamma}''=
\begin{bmatrix}
    \cos{\sigma}             & 0           & -\sin{\sigma} \\
    \sin{\sigma}\sin{\rho}   & \cos{\rho}  & \cos{\sigma}\sin{\rho} \\
    \sin{\sigma}\cos{\rho}   & -\sin{\rho} & \cos{\sigma}\cos{\rho}
\end{bmatrix},
\end{equation*}
where $\mathbf{\Gamma}'$ transforms vectors from $Oxyz$ to the body frame $Ox'y'z'$, and $\mathbf{\Gamma}''$ transforms  vectors from semi-orbital reference frame to $Oxyz$.

To write down the equations of motion we choose $\tau=n_{\Omega}t$ as independent variable.

\subsection{``Conservative evolution'' $(\mathbf{M}_{EC}=0$)}
The combined influence of the gravity gradient torque and the orbit evolution on the rotational motion of a satellite was studied in (\cite{C1972}, \cite{H1987}). In this case the magnitude of the angular momentum vector is an approximate integral of motion. Direction of $\mathbf{L}$ with respect to the semi-orbital frame is described by the equations:
\begin{equation}\label{eq:LH}
\frac{d\sigma}{d\tau}=\frac{\partial \mathcal{H}}{\partial p},\quad \frac{d p}{d\tau}=-\frac{\partial \mathcal{H}}{\partial\sigma},
\end{equation}
where
\begin{align}\label{eq:LHexpl}
 \mathcal{H} =& -\sqrt{1-p^2}\sin{\sigma}\sin{i}-p\cos{i}-\frac{\kappa_i p^2}{2\omega},\\
 p =& \cos{\rho},\quad \omega=\frac{\left|\mathbf{L}\right|}{A\omega_*},\quad \omega_*=\frac{3}{4}\left(1-\frac{C}{A}\right)\frac{\omega_o^2}{n_{\Omega}\kappa_i}, \nonumber \\
 \kappa_i =&\left(\cos^{2/3}{i}+\sin^{2/3}{i}\right)^{3/2}.	\nonumber
\end{align}

The dimensionless variable $\omega$ in \eqref{eq:LHexpl} denotes the ratio of the current angular velocity and $\omega_*$. For typical SSO $\omega_* \sim 100^\circ/s$, which is greater than observed angular velocities of rocket bodies immediately after separation. Therefore, without loss of generality, the dimensionless angular velocity $\omega$ will be assumed in our study
to be less then unity.

\begin{figure}[htb]
\center{\includegraphics[width=80mm]{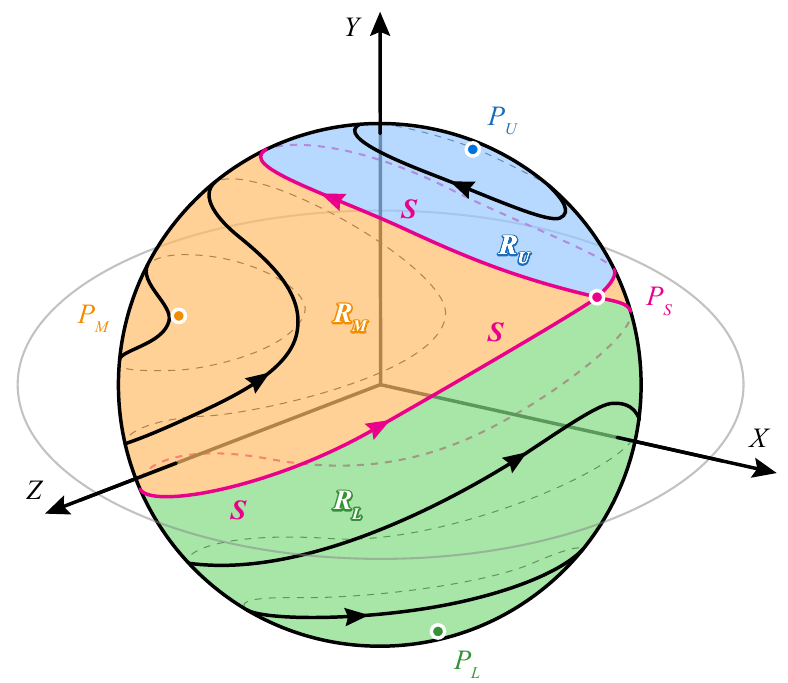}}
\caption{Cassini states and Cassini cycles}
\label{fig:CCycles}
\end{figure}

Equations \eqref{eq:LH} have stationary solutions, which are referred to as Cassini states (\cite{H1987}). It can be shown that for
\begin{equation}
\omega < 1
\label{eq:4states}
\end{equation}
there exist four Cassini states: three stable and one unstable (Fig.~\ref{fig:CCycles}).

If we draw trajectories of the unit vector
\[
\mathbf{e}_L=\left(\sqrt{1-p^2} \sin{\sigma},p,\sqrt{1-p^2}\cos{\sigma}\right)^T
\]
on the surface of a sphere $S^2$, the separatrices proceeding out of the unstable equilibrium divide this surface into three regions (Fig.~\ref{fig:CCycles}). Depending on positions of these regions with respect to the orbital plane, we shall denote them by $R_U$ (upper), $R_M$ (middle), and $R_L$ (lower). The stable Cassini states belonging to these regions are denoted by $P_U$, $P_M$, and $P_L$ respectively. The unstable Cassini state is denoted by $P_S$. The values of $p$ in the Cassini states are roots of the equation:
\begin{equation}
p^4 + 2\left(\frac{\omega}{\kappa_i}\right)\cos{i}\cdot p^3 +\left[\left(\frac{\omega}{\kappa_i}\right)^2 - 1\right]\cdot p^2 - 2 \left(\frac{\omega}{\kappa_i}\right) \cos{i} \cdot p - \left(\frac{\omega}{\kappa_i}\right)^2 \cos^2{i}=0.
\label{eq:CasEqw}
\end{equation}

For nearly polar retrograde orbits approximate expressions for the roots of the equation~\eqref{eq:CasEqw} can be easily obtained as:
\begin{align*}
P_U \text{ state:}\quad p&=\left(1-\omega^2\right)^{1/2} + O(\cos{i}), \quad \sigma = \frac{\pi}{2}; \\
P_S \text{ state:}\quad p&=\frac{\cos{i}}{1-\omega} + O(\cos^2{i}), \quad \sigma = \frac{\pi}{2}; \\
P_M \text{ state:}\quad p&=-\frac{\cos{i}}{1+\omega}+ O(\cos^2{i}), \quad \sigma = \frac{3\pi}{2}; \\
P_L \text{ state:}\quad p&=-\left(1-\omega^2\right)^{1/2} + O(\cos{i}), \quad \sigma = \frac{\pi}{2}.
\end{align*}

Let us use the value $h$ of Hamiltonian $\mathcal{H}$ along the corresponding solution and the value of the dimensionless angular velocity $\omega$ as parameters in the solution family of~\eqref{eq:LH}:
\begin{equation}
\sigma(\tau,h,\omega),~p(\tau,h,\omega).
\label{eq:SigmaP}
\end{equation}

We shall refer to periodical solutions of \eqref{eq:SigmaP} as \emph{Cassini cycles}. They are represented by closed curves on the sphere $S^2$ around the stable Cassini states (Fig.~\ref{fig:CCycles}).

\begin{figure}[htb]
\center{\includegraphics[width=122mm]{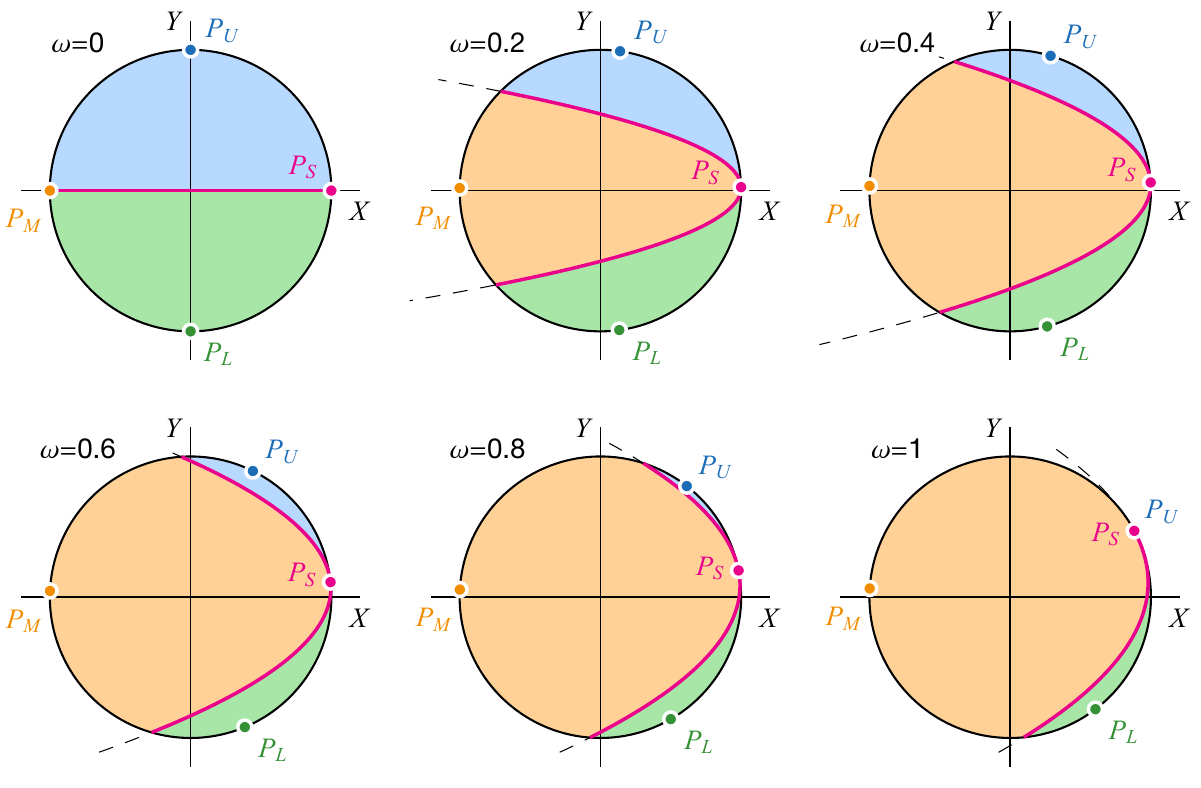}}
\caption{Cassini cycles and separatrices dividing the regions $R_U$, $R_M$, and $R_L$ for different values of dimensionless angular velocity $\omega$. Cases $\omega = 0$ and $\omega = 1$ are degenerate. At $\omega = 1$ Cassini states $P_U$ and $P_S$ merge, and $R_U$ region vanishes. For $\omega \rightarrow 0$ the width of $R_M$ region tends to zero.}
\label{fig:OmgCycles}
\end{figure}

Let us consider the values that the Hamiltonian $h(\omega)$ can take on the solutions of \eqref{eq:SigmaP}. The maximum $h_M(\omega)$ and the minimum $h_L(\omega)$ values of the Hamiltonian correspond to the stationary solutions $P_M$ and $P_L$, respectively. In the region $R_U$ the minimum of the Hamiltonian $h_U(\omega)$ is reached on the stationary solution $P_U$. Separatrices have the same value of the Hamiltonian $h=h_S(\omega)$ with the unstable stationary solution $P_S$. It follows that for the trajectories enclosing $P_M$ (i.e. trajectories belonging to $R_M$) hamiltonian value $h \in \left(h_S,h_M\right)$, for trajectories in $R_U$ hamiltonian value $h \in \left(h_U,h_S\right)$, and $h \in \left(h_L,h_S\right)$ for trajectories in $R_L$. Transformations of the regions $R_M$, $R_U$, and $R_L$ for different values of $\omega$ are shown in~Fig.~\ref{fig:OmgCycles}.


Vector $\mathbf{L}$ moves along a Cassini cycle with a period, which is calculated as follows
\begin{equation}
T_{Cassini}(h,\omega)=2\int\limits^{p^{max}}_{p_{min}}\frac{dp}{\dot{p}}=2\int\limits^{p^{max}}_{p_{min}}\frac{dp}{\sqrt{R_4(p)}}=\frac{4\omega}{\kappa_i}I_0.
\label{eq:TCassini}
\end{equation}
Here $p^{max}$ and $p_{min}$ are maximum and minimum value of $p$ for a given cycle, and designation $I_k$ is used for integrals
\begin{equation}\label{eq:Ik}
I_k=\int\limits^{p^{max}}_{p_{min}}\frac{p^k dp}{\sqrt{-(p-p_1)(p-p_2)(p-p_3)(p-p_4)}},
\end{equation}
where $p_1,...,p_4$ are roots of the equation $R_4(p)=0$,
\begin{equation}
R_4(p)=\left(1-p^2\right)\sin^2{i}-\left(h+p\cos{i}+\frac{\kappa_i p^2}{2\omega}\right)^2.
\label{eq:R4p}
\end{equation}

For values of $h\in(h_U,h_S)$, which correspond to Cassini cycles in $R_L$ or $R_U$, roots $p_1,...,p_4 \in \mathbb{R}^1$; for $h<h_U$ (Cassini cycle in $R_L$) or $h>h_S$ (Cassini cycle in $R_M$) roots $p_1,p_2 \in \mathbb{R}^1$, $p_3,p_4 \in \mathbb{C}^1$ $(p_4=\overline{p}_3)$. The values of integration limits in \eqref{eq:TCassini} for cycles in $R_L$ and $R_M$ are $p_{min}=p_1$, $p^{max}=p_2$; the corresponding values for cycles in $R_U$ are $p_{min}=p_3$, $p^{max}=p_4$ (rational roots of \eqref{eq:R4p} are arranged in ascending order of magnitude).

Analytic expressions for $I_k$ are given in Appendix.

\subsection{Derivation of evolution equations describing the eddy currents torque impact: averaging along the orbital motion and rotation about the center of mass}
Let us introduce the dimensionless torque due to eddy currents:
\begin{equation}
\mathbcal{M}_{EC} = (B_*^2S_*\omega_*)^{-1}\mathbf{M}_{EC}.
\end{equation}
where $B_*=\mu_0\mu_E/\left(4\pi R_O^3\right)$ is the characteristic magnitude of the magnetic field along the orbit, $S_*$ is the characteristic value of the magnetic tensor components (it is supposed that in the body frame $\mathbf{S}=S_*\mathbf{\Sigma}'$, $\mathbf{\Sigma}' =  \text{ diag}(1,1,\lambda)$). Let us denote by $\mathbcal{B}=\mathbf{B}/B_*$ the dimensionless vector of the magnetic field, whose components in the semi-orbital frame are given by:
\begin{equation*}
\mathcal{B}_X = (1-3\sin^2{u})\sin{i}, \quad \mathcal{B}_Y = \cos{i}, \quad \mathcal{B}_Z = -3\sin{u}\cos{u}\sin{i}.
\end{equation*}

To describe the evolution of rotation accounting for $\mathbf{M}_{EC}$ effect, we introduce the averaged equations analogous to \eqref{eq:LH}:

\begin{equation}\label{eq:HMec}
\frac{d\sigma}{d\tau}=\frac{\partial \mathcal{H}}{\partial p} +\varepsilon f_{\sigma},\quad \frac{d p}{d\tau}=-\frac{\partial \mathcal{H}}{\partial\sigma} + \varepsilon f_p, \quad \frac{d\omega}{dt}=\varepsilon f_{\omega},
\end{equation}
where
\begin{align*}
\varepsilon &= \frac{B_*^2S_*}{A n_{\Omega}}, \quad f_{\sigma}=\frac{1}{\omega\sqrt{1-p^2}}\left(\mathbf{e}_x,\left\langle \mathcal{M}_{EC}	\right\rangle_{\psi,u}\right), \\
f_p &= -\frac{1}{\omega}\sqrt{1-p^2}\left(\mathbf{e}_z,\left\langle \mathcal{M}_{EC}	\right\rangle_{\psi,u}\right), \quad f_{\omega} = \left(\mathbf{e}_L,\left\langle \mathcal{M}_{EC}	\right\rangle_{\psi,u}\right), \\
&\left\langle \cdot\right\rangle_{\psi,u} = \frac{1}{4\pi^2} \iint (\cdot) d \psi d u.
\end{align*}

To study the secular effects in the attitude motion with the use of the equations \eqref{eq:HMec}, we need to obtain an expression for the averaged dimensionless torque $\left\langle \mathbcal{M}_{EC} \right\rangle_{\psi,u}$. For convenience we shall represent the $\mathbcal{M}_{EC}$ as the sum of two terms, which will be averaged separately:
\begin{equation}
\label{MECpartition}
\mathbcal{M}_{EC}  = \mathbcal{M}_{EC,1} + \mathbcal{M}_{EC,2}.
\end{equation}
The term
\begin{equation}
\label{eq:MECdiss}
\mathbcal{M}_{EC,1} = \mathbcal{B} \times \boldsymbol{\Sigma} \left(\mathbcal{B}\times \omega \mathbf{e}_L \right)
\end{equation}
can be called a dissipative component, as it causes the slowing down of the object's rotation (\cite{O1967}). The second term is due to the change of magnetic field as the object moves along its orbit:
\begin{equation}
\label{eq:MECorb}
\mathbcal{M}_{EC,2} = \chi \mathbcal{B} \times \boldsymbol{\Sigma} \frac{d\mathbcal{B}}{du}= -\chi \mathbcal{B} \times \boldsymbol{\Sigma} \left(2\mathbcal{B} \times \mathbf{e}_Y + \sin{i}\cdot \mathbf{e}_Z\right), \quad \chi = \frac{\omega_D}{\omega_*}.
\end{equation}

\emph{Remark}: It follows from $\omega_D \ll \omega_*$ that $\chi \ll 1$ and $|\mathbcal{M}_{EC,1}| \gg |\mathbcal{M}_{EC,2}|$
at the stage of fast rotation $(\omega \sim 1)$. Nevertheless, our numeric experiments show that if the influence of $\mathbcal{M}_{EC,2}$ is neglected, there appears a significant discrepancy between the solutions of non-averaged equations and solutions of \eqref{eq:HMec}, which arises long before the moment when the decelerated angular velocity value becomes comparable to $\omega_D$.

Let us start the averaging procedure with the first term of $\mathbcal{M}_{EC}$. Introducing for vector $\mathbcal{B}$ the matrix
\[
\hat{\mathbcal{B}}=
\begin{bmatrix}
    0               & -\mathcal{B}_Z & \mathcal{B}_Y \\
    \mathcal{B}_Z   & 0              & -\mathcal{B}_X \\
    -\mathcal{B}_Y  & \mathcal{B}_X  & 0
\end{bmatrix},
\]
we shall transform the expression for $\mathbcal{M}_{EC,1}$ as follows:
\begin{equation}
\mathbcal{M}_{EC,1} = \omega \hat{\mathbcal{B}} \boldsymbol{\Sigma} \hat{\mathbcal{B}} \mathbf{e}_L = \omega \hat{\mathbcal{B}}\mathbf{{\Gamma}''}^T\mathbf{{\Gamma}'}^T \boldsymbol{\Sigma}' \mathbf{\Gamma}'\mathbf{\Gamma}''\hat{\mathbcal{B}} \mathbf{e}_L.
\label{eq:MEC1}
\end{equation}

Supposing that the components of magnetic field vector are written in the semi-orbital reference frame, we average the expressions~\eqref{eq:MEC1} over $\psi$:
\[
\left\langle \mathbcal{M}_{EC,1}	\right\rangle_{\psi} = \omega \hat{\mathbcal{B}} \left\langle \boldsymbol{\Sigma}	\right\rangle_{\psi} \hat{\mathbcal{B}} \mathbf{e}_L,
\]
where
\[
\left\langle \boldsymbol{\Sigma}	\right\rangle_{\psi} = \mathbf{{\Gamma}''}^T\left\langle \boldsymbol{\Sigma}''	\right\rangle_{\psi}\mathbf{\Gamma}'', \quad \left\langle \boldsymbol{\Sigma}''	\right\rangle_{\psi} = \left\langle\mathbf{{\Gamma}'}^T \boldsymbol{\Sigma}'	\mathbf{\Gamma}'\right\rangle_{\psi}=
\begin{bmatrix}
    \frac{1+\lambda}{2} & 0 & 0 \\
    0                   & 1 & 0 \\
    0                   & 0  & \frac{1+\lambda}{2}
\end{bmatrix}.
\]
Taking into account
\[
\mathbf{{\Gamma}''}^T\left\langle \boldsymbol{\Sigma}''	\right\rangle_{\psi}\mathbf{\Gamma}''=\frac{1+\lambda}{2}\mathbf{E}_3+\frac{1-\lambda}{2}\mathbf{e}_L\mathbf{e}_L^T,
\]
where $\mathbf{E}_3$ is the identity matrix, we obtain the following expression for $\left\langle \mathbcal{M}_{EC,1} \right\rangle_{\psi}$:
\begin{equation}
\left\langle \mathbcal{M}_{EC,1}	\right\rangle_{\psi}=\frac{\left(1+\lambda\right)\omega}{2}\hat{\mathbcal{B}}^2\mathbf{e}_L.
\label{eq:MEC1t}
\end{equation}
Averaging \eqref{eq:MEC1t} along the orbital motion yields:
\begin{equation}
\left\langle \mathbcal{M}_{EC,1}	\right\rangle_{\psi,u}=-\frac{\left(1+\lambda\right)\omega}{2}\boldsymbol{\Xi}\mathbf{e}_L,
\label{MEC1av}
\end{equation}
where
\begin{align*}
\boldsymbol{\Xi} &=
\begin{bmatrix}
    \xi_{11} & \xi_{12} & 0 \\
    \xi_{21} & \xi_{22} & 0 \\
    0        & 0  & \xi_{33}
\end{bmatrix}, \\
\xi_{11}&=1+\frac{1}{8}\sin^2{i}, \quad \xi_{12}=\xi_{21}=\frac{1}{4}\sin{2i}, \\
\xi_{22}&=\frac{5}{2}\sin^2{i}, \quad \xi_{33}=1+\frac{3}{8}\sin^2{i}.
\end{align*}

If $\sin{i} \neq 0$ matrix $\boldsymbol{\Xi}$ is positive definite. The greatest and the smallest eigenvalues of this matrix are
\[
\eta_{min,max}=\frac{1}{2}\left(1+\frac{21}{8}\sin^2{i}\pm \sqrt{1-\frac{15}{4}\sin^2{i}+\frac{297}{64}\sin^4{i}}\right).
\]

Let us proceed to the averaging of the second term of the $\mathbcal{M}_{EC}$ torque. In the expression for $\mathbcal{M}_{EC,2}$ we shall also replace the dimensionless vector $\mathbcal{B}$ by the matrix $\hat{\mathbcal{B}}$ :
\begin{equation}
\label{eq:MEC2}
\mathbcal{M}_{EC,2}=-\chi\left[2\hat{\mathbcal{B}}\boldsymbol{\Sigma} \hat{\mathbcal{B}} \mathbf{e}_Y + \sin{i}\cdot\hat{\mathbcal{B}}\boldsymbol{\Sigma} \mathbf{e}_Z \right].
\end{equation}

Averaging \eqref{eq:MEC2} yields:
\[
\left\langle\mathbcal{M}_{EC,2}\right\rangle_{\psi,u}=-\chi\left[2\left\langle\hat{\mathbcal{B}}\left\langle\boldsymbol{\Sigma}\right\rangle_{\psi}\hat{\mathbcal{B}}\right\rangle_u\mathbf{e}_Y+\sin{i}\left\langle\hat{\mathbcal{B}}\right\rangle_u\left\langle\boldsymbol{\Sigma}\right\rangle_{\psi}\mathbf{e}_Z\right].
\]

The following relations are satisfied:
\begin{align}\label{eq:MEC2parts}
2\left\langle\hat{\mathbcal{B}}\left\langle\boldsymbol{\Sigma}\right\rangle_{\psi}\hat{\mathbcal{B}}\right\rangle_u\mathbf{e}_Y=&-(1+\lambda)(\xi_{12}\mathbf{e}_X+\xi_{22}\mathbf{e}_Y)-\\
&-(1-\lambda)\sin{\rho}\left[\left\langle\mathbcal{B}_x\mathbcal{B}_z\right\rangle_u\mathbf{e}_x-\left\langle\mathbcal{B}^2_x\right\rangle_u\mathbf{e}_z\right],\nonumber \\
\left\langle\hat{\mathbcal{B}}\right\rangle_u\left\langle\boldsymbol{\Sigma}\right\rangle_{\psi}\mathbf{e}_Z =&\frac{1}{2}\left\{(1+\lambda)\left[\left\langle\mathbcal{B}_Y\right\rangle_u\mathbf{e}_X-\left\langle\mathbcal{B}_X\right\rangle_u\mathbf{e}_Y\right]\right.+ \nonumber \\ &\left. +(1-\lambda)\sin{\rho}\cos{\sigma}\left[-\left\langle\mathbcal{B}_z\right\rangle_u\mathbf{e}_x+\left\langle\mathbcal{B}_x\right\rangle_u\mathbf{e}_z\right]\right\}, \nonumber
\end{align}
where
\begin{gather*}
\left\langle\mathcal{B}_X\right\rangle_u=-\frac{1}{2}\sin{i},\quad \left\langle\mathcal{B}_Y\right\rangle_u=\cos{i}, \\
\left\langle\mathcal{B}_x\right\rangle_u=-\frac{1}{2}\cos{\sigma}\sin{i},\quad \left\langle\mathcal{B}_z\right\rangle_u=-\frac{1}{2}\sin{i}\sin{\sigma}\cos{\rho}-\cos{i}\sin{\rho}, \\
\left\langle\mathcal{B}^2_x\right\rangle_u=\frac{1}{4}\sin^2{i}\left(\frac{9}{2}+\cos^2{\sigma}\right),\quad \left\langle\mathcal{B}_x\mathcal{B}_z\right\rangle_u=\frac{1}{2}\sin{i}\cos{\sigma}\left(\sin{\rho}\cos{i}+\frac{1}{2}\sin{i}\cos{\rho}\sin{\sigma}\right).
\end{gather*}

Using the relations~\eqref{eq:MEC2parts}, we obtain:
\begin{equation}\label{MEC2av}
\left\langle\mathbcal{M}_{EC,2}\right\rangle_{\psi,u}=\frac{9}{8}\chi\sin^2{i}\left[2(1+\lambda)\mathbf{e}_Y+\sin{\rho}(\lambda-1)\mathbf{e}_z\right].
\end{equation}

The equations~\eqref{eq:HMec} are of instrumental value for us. We shall use them to construct evolution equations, describing the rotational motion of the object at long time intervals. It may be difficult to draw definite conclusions about the properties of motion directly from the equations~\eqref{eq:HMec}. However, it is worthwhile noticing that the last equation in the system~\eqref{eq:HMec} allows writing down the following inequalities, characterizing the changes in value of the dimensionless angular velocity during the object's fast rotation (\cite{S1982}):
\begin{equation}
exp\left[-\frac{\varepsilon \eta_{max}(1+\lambda)(\tau-\tau_0)}{2}\right]\leq \omega(\tau)\leq exp\left[-\frac{\varepsilon \eta_{min}(1+\lambda)(\tau-\tau_0)}{2}\right]
\label{eq:omegaineq}
\end{equation}

Inequalities~\eqref{eq:omegaineq} become invalid when the magnitude of angular velocity becomes comparable to $\omega_D$.
	
\subsection{Averaging along Cassini cycles}\label{subsec:avcas}
For small $\varepsilon$ the behavior of variables $\sigma$ and $p$ in solutions of the system~\eqref{eq:HMec} can be described as a Cassini cycle with slowly changing parameters $h, \omega$. Let us write the equations for $h, \omega$ and average them along the solutions of \eqref{eq:SigmaP}:
\begin{align}\label{eq:HW}
\frac{dh}{dt}=&\frac{\varepsilon}{T_{Cassini}(h,\omega)}\int\limits_0^{T_{Cassini}}{\left[\frac{\partial\mathcal{H}}{\partial\sigma}f_{\sigma}(\sigma(\tau,h,\omega),p(\tau,h,\omega),\omega)\right.}+\\&+\left.\frac{\partial\mathcal{H}}{\partial p}f_{p}(\sigma(\tau,h,\omega),p(\tau,h,\omega),\omega)+\frac{\partial\mathcal{H}}{\partial \omega}f_{\omega}(\sigma(\tau,h,\omega),p(\tau,h,\omega),\omega)\right]d\tau\nonumber,\\
\frac{d\omega}{dt}=&\frac{\varepsilon}{T_{Cassini}(h,\omega)}\int\limits_0^{T_{Cassini}}{f_{\omega}(\sigma(\tau,h,\omega),p(\tau,h,\omega),\omega)}d\tau. \nonumber
\end{align}

We shall refer to \eqref{eq:HW} as evolution equations. For convenience let us write the right-hand sides of equations~\eqref{eq:HW} as sums of integrals $I_k$, which were introduced previously by the equation \eqref{eq:Ik}:
\begin{equation}
\frac{dh}{dt}=\frac{\varepsilon}{2I_0}\sum\limits_{k=0}^{6}{C_k I_k}, \quad \frac{d\omega}{dt}=\frac{\varepsilon}{2I_0}\sum\limits_{k=0}^{4}{W_k I_k},
\label{eq:HWint}
\end{equation}
where
\begin{align*}
C_k&=-(1+\lambda)c^{(1)}_k +\frac{9}{4}\chi \sin^2{i}\cdot c^{(2)}_k, \quad W_k= -(1+\lambda)w^{(1)}_k +\frac{9}{4}\chi \sin^2{i}\cdot w^{(2)}_k, \\
c^{(1)}_0&=\frac{1}{8}h(1+2h^2+3\cos{2i}), \quad c^{(1)}_1=\frac{3}{16}\cos{i}\left[8h\left(\frac{\kappa_i}{3\omega}+h\right)+9\cos{2i}-1\right],\\
c^{(1)}_2&=\frac{9}{32\omega}\left[2h\omega+3\kappa_i+(5\kappa_i+6h\omega)\cos{2i}\right],\quad c^{(1)}_3=\frac{\kappa_i}{4\omega^2}(\kappa_i-3h\omega)\cos{i},\\
c^{(1)}_4&=-\frac{3\kappa_i}{16\omega^2}\left[h\kappa_i+3\omega(1+3\cos{2i})\right], \quad c^{(1)}_5=-\frac{3\kappa_i^2}{4\omega^2}\cos{i}, \quad c^{(1)}_6=-\frac{\kappa_i^3}{16\omega^3},\\
c^{(2)}_0&=-\frac{3+\lambda}{\omega}\cos{i}, \quad c^{(2)}_1=-\frac{3+\lambda}{\omega^2}(\kappa_i +h\omega), \quad c^{(2)}_3=\frac{5+3\lambda}{2\omega^2}\kappa_i, \\
w^{(1)}_0&=\frac{\omega}{16}(19-4h^2-3\cos{2i}), \quad w^{(1)}_1=-\frac{3}{2}h\omega\cos{i}, \\
w^{(1)}_2&=-\frac{1}{16}\left[4h\kappa_i+9\omega(1+3\cos{2i})\right], \quad w^{(1)}_3=-\frac{3}{4}\kappa_i\cos{i}, \quad w^{(1)}_4 =-\frac{\kappa_i^2}{16\omega}, \\
w^{(2)}_1 &= 2(1+\lambda).
\end{align*}
Upper index in $c^{(i)}_k$, $w^{(i)}_k$ denotes the corresponding component of the torque due to eddy currents \eqref{MECpartition}. All coefficients that are not listed here equal zero.

\emph{Remark}:
In general case an object may be asymmetrical. Let us denote its principle moments of inertia by $A'$, $B'$, $C'$ ($A'\ge B'\ge C'$, $A>C$) and by $S_{x'x'}'$, $S_{y'y'}'$, $S_{z'z'}'$ the diagonal components of its magnetic tensor written in the principal axes of inertia (magnetic tensor itself in these axes does not have to be diagonal). Using the evolution equations~\eqref{eq:HWint} to study the secular effects in the attitude dynamics of such object, requires the ``effective'' parameters $A$, $C$, $S_{x'x'}$, $S_{z'z'}$, which are calculated as follows:
\begin{equation}\label{eq:ParGen}
\begin{gathered}
A=A', \quad C=C'+B'-A', \\
S_{x'x'}=S_{x'x'}', \quad S_{z'z'}=S_{z'z'}'+S_{y'y'}'-S_{x'x'}'.
\end{gathered}
\end{equation}
These effective parameters are then used to calculate the values of all the auxiliary quantities in~\eqref{eq:HWint}.

\subsection{Evolution equations and qualitative analysis of large debris objects' dynamics for fast rotations about the center of mass}\label{subsec:Rs}

For better understanding of the rotational motion evolution, let us draw phase portraits for the system~\eqref{eq:HW}. In order to see how far a solution goes into one of the regions $R_L$--$R_U$, we shall use ``relative'' variables $\tilde{h}$ instead of $h$:
\begin{equation}\label{hnormalization}
\begin{gathered}
\tilde{h}=\frac{h-h_L(\omega)}{h_S(\omega)-h_L(\omega)} \text{ (in $R_L$ region)}, \quad \tilde{h}=\frac{h-h_S(\omega)}{h_M(\omega)-h_S(\omega)} \text{ (in $R_M$ region)},\\
\tilde{h}=\frac{h-h_U(\omega)}{h_S(\omega)-h_U(\omega)} \text{ (in $R_U$ region)}.
\end{gathered}
\end{equation}

We shall also use the auxiliary value $\tilde{\omega}$ of angular velocity of the object nondimentionalized by mean motion $\omega_o$. It is related to the previously introduced $\omega_*$ by the formula:
\begin{equation}\label{omegawave}
\tilde{\omega}= \frac{\omega_*}{\omega_o} \omega.
\end{equation}

\begin{figure}[htb]
\center{\includegraphics[width=61mm]{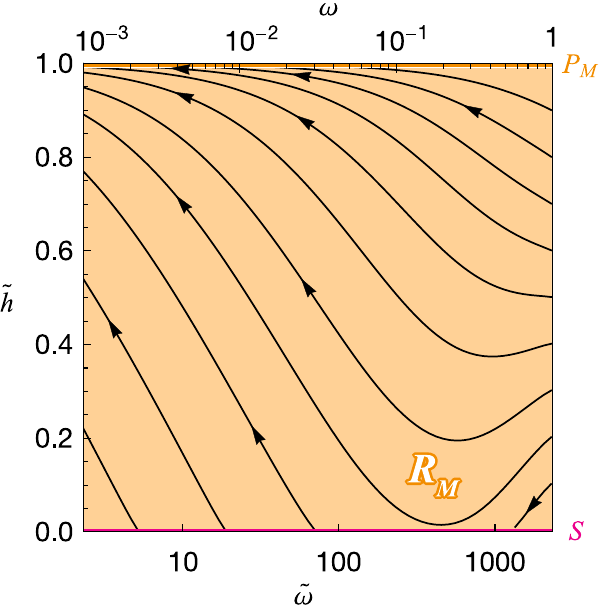}}
\center{\includegraphics[width=61mm]{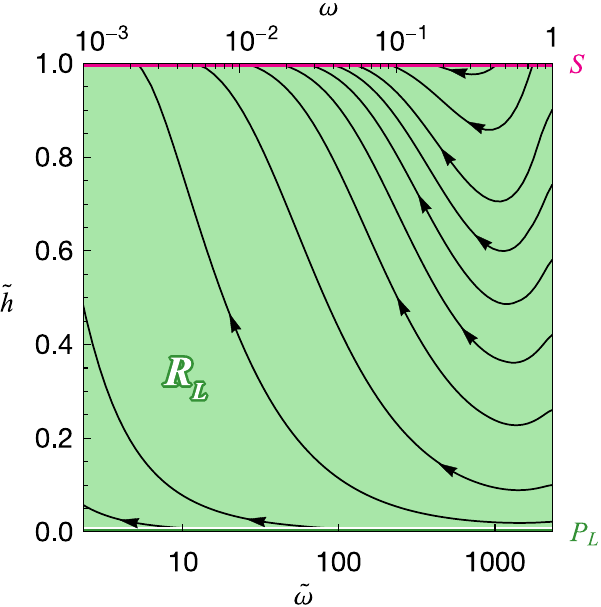}\includegraphics[width=61mm]{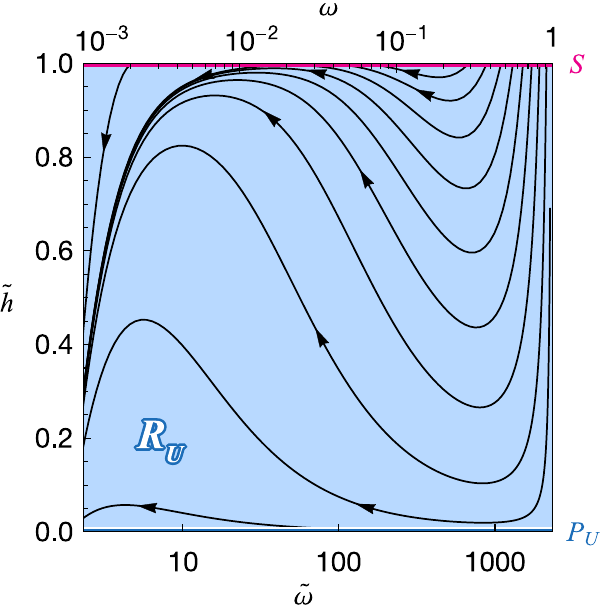}}
\caption{Evolution of ``osculating'' Cassini cycles' parameters}
\label{fig:oscCC}
\end{figure}

Figure~\ref{fig:oscCC} shows the phase portraits in the space $\left(\tilde{\omega},\tilde{h}\right)$, which describe the long-term evolution of Cassini cycles.
The interval of angular velocities here corresponds to the applicability range of the averaged equations \eqref{eq:HWint}, i.e. from angular velocities comparable to mean motion ($\tilde\omega\sim1$) to critical angular velocity value, at which two Cassini states vanish ($\omega=1$). As the real values of angular velocity are usually much smaller than $\omega_*$, this practically covers all possible variants of the exponential deceleration stage.
Generally, trajectories of the system depend on the object's parameters (the depicted case corresponds to the set listed in Table~\ref{tbl:params}). However, this dependence is weak and Fig.~\ref{fig:oscCC} correctly reflects the qualitative evolution of Cassini cycles in SSO for most objects.

Let us analyze the acquired results. Because of exponential deceleration due to eddy currents torque, all trajectories head towards the zone of lower angular velocity values (Fig.~\ref{fig:oscCC}). Map of $R_M$ reveals that most of the trajectories starting in this region do not leave $R_M$ and converge towards the Cassini state $P_M$. The trajectories entering this region through the separatrices $S$ also tend to $P_M$. Most of the trajectories in $R_L$ are directed towards the separatrix and cross it, leaving $R_L$.

\begin{figure}[htb]
\center{\includegraphics[width=80mm]{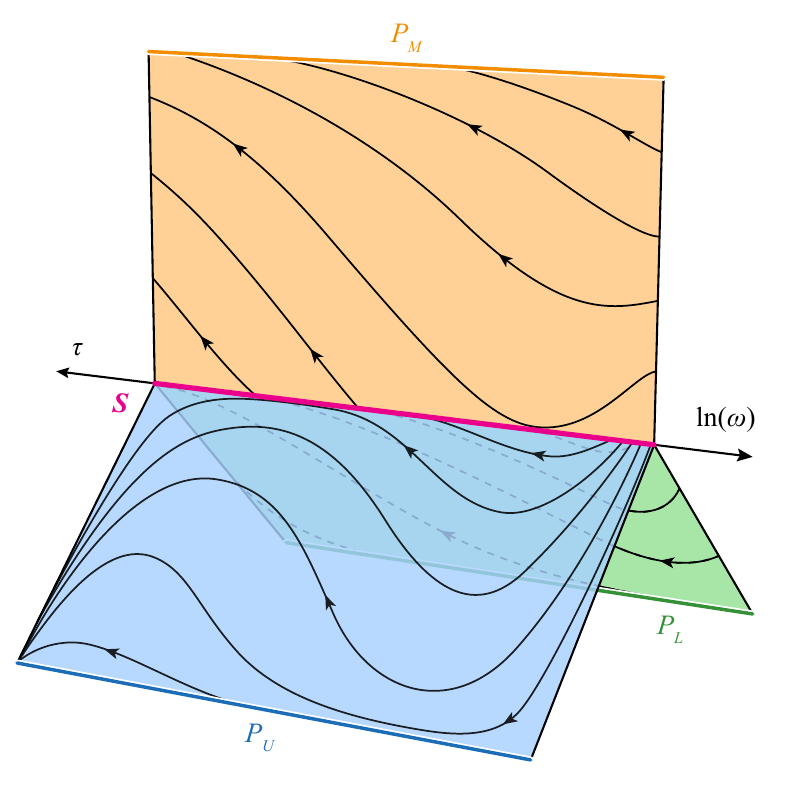}}
\caption{Multi-sheet phase portrait: phase portraits for regions $R_U$, $R_M$, and $R_L$ put together}
\label{fig:MSheet}
\end{figure}

The dynamics in the region $R_U$ is most interesting. Typical trajectories in this region are ``S''-shaped. The downward flow of trajectories in the region of high angular velocities ($\tilde{\omega}\gtrsim 700$) exists mainly as an artifact of normalization \eqref{hnormalization}. Because at $\omega=1$ region $R_U$ vanishes and point $P_U$ merges with the separatrix (Fig.~\ref{fig:OmgCycles}) for $\omega\lesssim 1$ the apparent general direction of trajectories in Figure~\ref{fig:MSheet} is defined by the rapid inflation of $R_U$. The dynamics in the rest of the region $R_U$ is characterized by the change in trajectories' flow direction from upward to downward at $\tilde{\omega}\sim10\div50$. It is governed by the interplay of two components in eddy currents torque: dissipative $\mathbcal{M}_{EC,1}$ and orbital $\mathbcal{M}_{EC,2}$ given by \eqref{eq:MECdiss} and \eqref{eq:MECorb} respectively.

As $|\mathbcal{M}_{EC,1}|\propto \omega$ and $\mathbcal{M}_{EC,2}$ does not depend on $\omega$, the evolution for very fast spins is defined by the dissipative component of the eddy currents torque, which drives the angular momentum towards the orbital plane. Consequently for $\tilde{\omega}\in (50 ,700)$ the flows of trajectories in the $R_L$ and $R_U$ regions are directed towards the separatrices and look very much alike.

The orbital component of the eddy currents torque, as seen from \eqref{MEC2av}, has a part directed along $\mathbf{e}_Y$, which is close to direction towards $P_U$ (Fig.~\ref{fig:OmgCycles}). Therefore, this component spins the debris object up about the orbital normal and results in deflection of trajectories in $R_U$ towards $P_U$ at $\tilde{\omega}\sim10\div50$. It should be noted, that this interval corresponds to relatively fast spins for which $|\mathbcal{M}_{EC,1}| \gg |\mathbcal{M}_{EC,2}|$. However, near the separatrix the directions of these torques turn out to be such that $\mathbcal{M}_{EC,1}$ mainly affects angular velocity value, while the direction of rotational axis is primarily influenced by $\mathbcal{M}_{EC,2}$. Thus the orbital component of eddy currents torque starts to have a noticeable effect on attitude dynamics long before the value of the angular velocity becomes comparable to mean motion.

In other words, the orbital component of the eddy currents torque keeps most of the trajectories in $R_U$ from crossing the separatrix, while in $R_L$ it only increases the rate at which trajectories approach the separatrix.

To illustrate the transitions of phase trajectories between the regions, phase portraits (Fig.~\ref{fig:oscCC}) are joined together along the separatrices, as shown in Figure~\ref{fig:MSheet}. Directions of transitions are indicated in Table 2.

\noindent
\begin{table}[hb]
\begin{center}
\captionsetup{justification=centering}
\caption{Transitions between regions $R_L$, $R_M$, $R_U$ through separatrices}\label{tbl:transitions}
\begin{tabular}{cccccc}
\hline \\
\vspace{5pt}
$\omega < \omega_{t1}$ & $\omega_{t1} < \omega < \omega_{t2}$ & $\omega_{t2} < \omega < \omega_{t3}$ & $\omega_{t3} < \omega < \omega_{t4}$ & $\omega_{t4} < \omega < 1$ \vspace{5pt}\\
\hline
$R_L {\nearrow\rule[-5pt]{0pt}{7pt}\atop \searrow}
{\displaystyle R_U \rule{0pt}{15pt}\atop \displaystyle R_M \rule{0pt}{15pt}}$\vspace{10pt}&
${\displaystyle R_U \rule{0pt}{15pt}\atop \displaystyle R_L \rule{0pt}{15pt}}
{\searrow\rule[-5pt]{0pt}{7pt}\atop \nearrow} R_M$ &
$R_L {\nearrow\rule[-5pt]{0pt}{7pt}\atop \searrow}
{\displaystyle R_U \rule{0pt}{15pt}\atop \displaystyle R_M \rule{0pt}{15pt}}$&
${\displaystyle R_M \rule{0pt}{15pt}\atop \displaystyle R_L \rule{0pt}{15pt}}
{\searrow\rule[-5pt]{0pt}{7pt}\atop \nearrow} R_U$ &
$R_M {\nearrow\rule[-5pt]{0pt}{7pt}\atop \searrow}
{\displaystyle R_U \rule{0pt}{15pt}\atop \displaystyle R_L \rule{0pt}{15pt}}$&
\\
\hline
\end{tabular}
\end{center}
\end{table}

\emph{Remark}:
$\omega_{t1} \approx 0.03$, $\omega_{t2} \approx 0.14$, $\omega_{t3} \approx 0.2$, and $\omega_{t4} \approx 0.27$ are the values of the dimensionless angular velocity which separate the attracting and repelling segments of the border $S$ of the phase portraits Fig.~\ref{fig:oscCC}. Transitions in the odd columns have quasi-probabilistic character.

Most of the trajectories starting in $R_M$ and $R_U$ remain in respective regions. In contrast to this, almost all trajectories from the region $R_L$ do cross the separatrix and transit to $R_M$ or $R_U$. This transition has a quasi-probabilistic nature and the probabilities of a trajectory going to either one of those regions depends on the object's parameters: the greater relative value of the eddy currents' torque leads to the greater probability of transition into $R_U$.

The middle region $R_M$ quickly becomes very narrow (Fig.~\ref{fig:OmgCycles}) because of the exponential deceleration of $\omega$. Thus, transition into region $R_M$ resembles a capture into oscillations about $P_M$, which for SSO roughly corresponds to the direction towards the south celestial pole. This transition can also be considered as a resonance phenomenon, since mean precession rate of the angular momentum vector in the inertial reference frame equals to the precession rate of the orbital plane ($\langle\dot{\sigma}\rangle =0$).

\subsection{Classification of long-term evolution scenarios: mapping the space of initial conditions}\label{subsection:map}

To study how the attitude motion evolution depends on the initial values of $\rho$ and $\sigma$ we consider an object with parameters given in Table~\ref{tbl:params} rotating with the angular velocity $12^\circ/s$ ($\tilde{\omega}=200$, $\omega\approx0.086$). This value of the initial angular velocity, on the one hand, is close to angular velocity of real rocket bodies after payload separation (\cite{depontieu1997}), and, on the other hand, corresponds to approximately even partition of the initial conditions space to regions $R_L$, $R_M$, and $R_U$ in terms of their area (Fig.~\ref{fig:OmgCycles}), thus producing a representative set of different dynamical cases.

\begin{figure}[htb]
\center{\includegraphics[width=122mm]{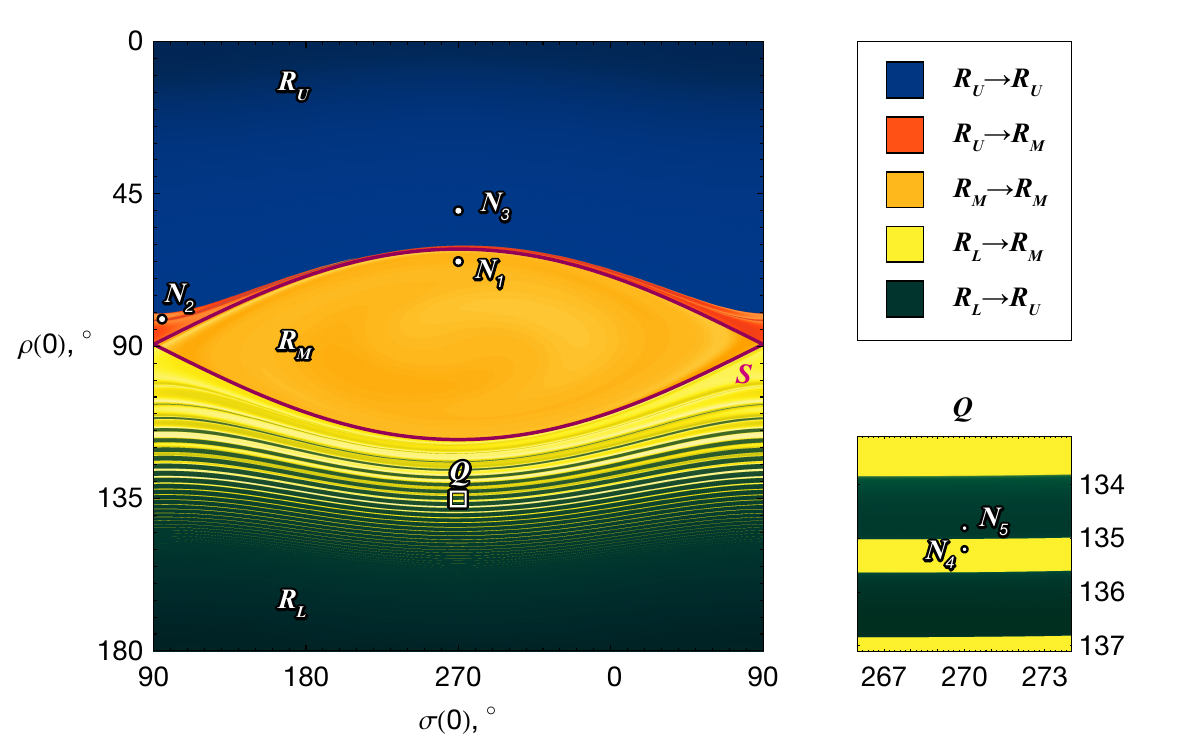}}
\caption{Map of different evolutional scenarios in the space of initial conditions for double-averaged system. Numbers denote initial conditions used in numerical simulations in section \ref{sec:numeric}. On the right side the enlarged sector $Q$ of the map is shown.}
\label{fig:Map}
\end{figure}

Let us classify different scenarios of the attitude motion long-term evolution according to pairs of regions $R_{i}\rightarrow R_{f}$, where index $i$ denotes the region in which the evolution starts, and index $f$ indicates the region in which the system is found by the end of the exponential decay stage. Both indices $i,f \in \{L,M,U\}$. This notation implies nine possible scenarios. However, judging by the phase portrait of the region $R_M$ (Fig.~\ref{fig:oscCC}) what starts in $R_M$ stays in $R_M$, and thus only one out of three $R_{M}\rightarrow R_{f}$ scenarios actually exists -- $R_{M}\rightarrow R_{M}$. Also there are no transitions leading into $R_L$, therefore $R_{U}\rightarrow R_{L}$ is impossible as well as $R_{M}\rightarrow R_{L}$. Lastly, the scenario $R_{L}\rightarrow R_{L}$, although feasible according to Figure~\ref{fig:oscCC}, turns out to have a negligibly small phase area of the corresponding initial conditions. Thus, to all practical purposes there remain five different scenarios of long-term evolution:
\[
R_U\rightarrow R_U, \quad R_U\rightarrow R_M, \quad R_M\rightarrow R_M, \quad R_L\rightarrow R_M, \quad R_L\rightarrow R_U.
\]
To give an idea of how the phase trajectories corresponding to different scenarios are mixed we  present in Fig.~\ref{fig:Map} the partition of the initial conditions ($\rho$ and $\sigma$) for the averaged system \eqref{eq:HMec}. It does not predict exactly the type of evolution in the original non-averaged system. Nevertheless, it correctly characterizes the sensitivity to variation of initial conditions.

According to the analysis in Section~\ref{subsec:Rs}, if evolution starts in $R_U$, the system in most cases remains in $R_U$ during the whole exponential decay stage. However, if initial conditions are close enough to separatrix, the transit $R_U\rightarrow R_M$ can take place, as revealed by the narrow band between $R_U$ and $R_M$ corresponding to this scenario (Fig.~\ref{fig:Map}).

In the region $R_L$ the domains corresponding to scenarios $R_{L}\rightarrow R_{M}$ and $R_{L}\rightarrow R_{U}$ take the form of tightly interleaved stripes. This is the reason behind the previously discussed quasi-probabilistic nature of these transitions. A small variation of initial $\rho$ value can lead to change of evolution scenario, therefore uncertainty in initial conditions, which always exists in practice, does not allow uniquely determining the type of subsequent evolution.

\section{Numerical study of fast rotation evolution of large space debris objects in SSO}\label{sec:numeric}
\subsection{Numerical Simulation Setup}
The regimes of motion described earlier in Section~\ref{sec:analytic} are of temporary character. They are destroyed when the angular rate decreases to become comparable to $\omega_o$. From this point the rotation evolution cannot be described by the equations~\eqref{eq:HW}, derived under the assumption of the object's fast rotation. Hence to study the transformations of the motion regimes and discover the final motion modes, we carried out numerical experiments. Furthermore, the numeric simulation was necessary to corroborate the conclusions drawn in Section~\ref{sec:analytic} for objects with realistic parameter values (listed in Table~\ref{tbl:params}), because in that case $\varepsilon \approx 0.5$. Yet, even for such values of parameter $\varepsilon$ the averaged equations \eqref{eq:HW} proved to be accurate enough to describe both qualitative and quantitative properties of the object's motion in the stage of exponential decay.

In all simulations the following motion characteristics were kept track of:
\begin{itemize}
	\item absolute value of angular velocity;
	\item angle $\delta$ between the axis with the least moment of inertia and the local vertical;
	\item angles $\rho$ and $\sigma$, describing the angular momentum $\mathbf{L}$ direction.
\end{itemize}

All simulations start from an orbital position corresponding to the crossing of the ascending node, and with the angular momentum directed along the axis with the greatest moment of inertia, i.e. ``flat'' spin. The initial value of angular velocity equals $200\omega_o$ and is the same as in Section~\ref{subsection:map}.


\subsection{Simulation results: validation of the averaged equations \eqref{eq:HMec} and \eqref{eq:HWint}}\label{sec:validation}
\begin{figure}[htb]
\center{\includegraphics[width=122mm]{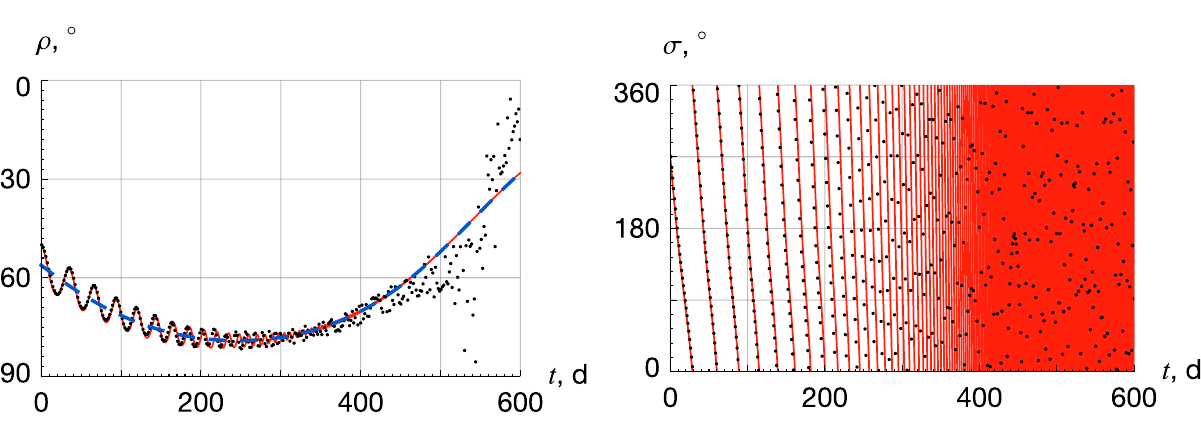}}
\caption{Comparison of numerical simulation results (black points) with solutions of double-averaged system (solid red line) and thrice averaged system (dashed blue line on $\rho(t)$ plot)}
\label{fig:valid}
\end{figure}

Figure~\ref{fig:valid} shows the comparison of numerical simulation results with the solutions of double averaged system \eqref{eq:HMec} and thrice averaged system \eqref{eq:HWint}. For the latter one we used the mean value of angle $\rho$ in the Cassini cycle for any given $h$. Thrice averaged system is not presented on $\sigma(t)$ plot, because this angle defines the position on the cycle, and this information vanishes when the motion is averaged along Cassini cycles. For both averaged systems the set of parameters modified in accordance with \eqref{eq:ParGen} was used.

It can be seen, that solution of \eqref{eq:HMec} (drawn by the red line) closely follows the numerical results (black points) at the beginning, but starts slightly deviate from them as time grows. This happens due to rise of fluctuations, caused by gravity gradient torque at smaller angular velocities. The conformity is totally lost at time $t\approx500 d$ with the end of exponential decay and the beginning of slow chaos stage. Angular velocity value at this moment equals approximately $4\omega_o$.

Figure~\ref{fig:valid} also demonstrates how accurately the solution of the thrice averaged system (blue line) describes the secular evolution of the angle $\rho$.

\subsection{Simulation results: exponential deceleration and slow chaotic stabilization}

\begin{figure}[htb]
\center{\includegraphics[width=122mm]{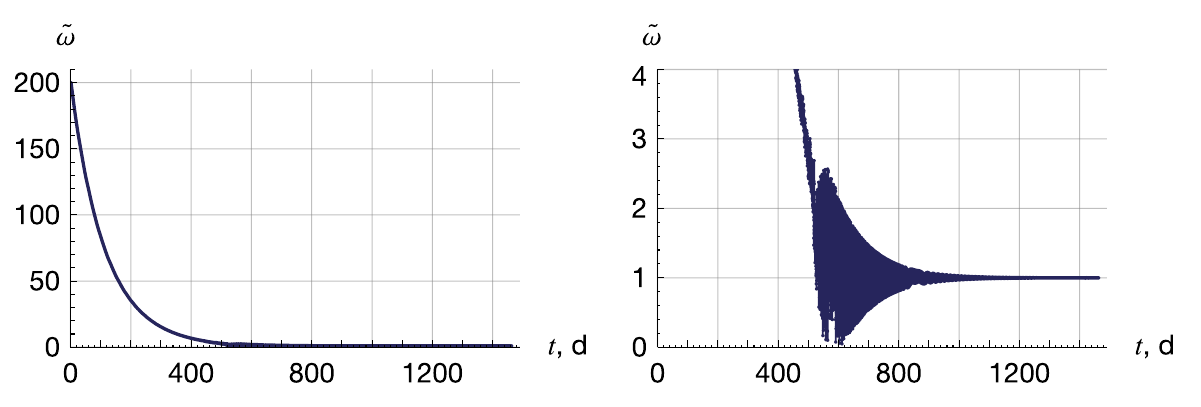}}
\caption{Evolution of angular velocity to mean motion ratio (a), scaled fragment of evolution showing the gravitational capture stage in greater detail (b)}
\label{fig:DeCap}
\end{figure}

Numerical experiments show that direction of initial angular velocity has no significant effect on subsequent behavior of its absolute value. Typical dependence of angular velocity on time is presented in Figure~\ref{fig:DeCap}. It takes about $500\div 600$ days for angular velocity to decrease to values comparable to the mean motion. The right side of Figure~\ref{fig:DeCap} shows scaled graph so as to demonstrate the stage of slow chaotic motion preceding the gravitational stabilization of the object $(\tilde{\omega}\sim 1)$.

\begin{figure}[htb]
\center{\includegraphics[width=122mm]{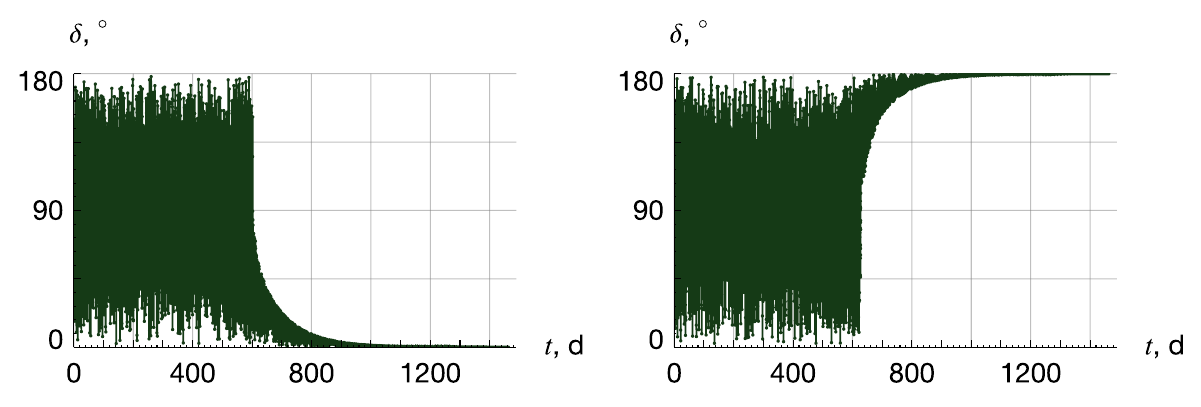}}
\caption{Two different variants of gravitational stabilization}
\label{fig:GravCapture}
\end{figure}

The moment $t_{G}$ when the graph $\delta(t)$ crosses the line $\delta=90^\circ$ for the last time is natural to define as moment of gravitational capture. It is clearly seen in Figure~\ref{fig:GravCapture} and $t_{G}=600\div 640$ days for different simulations. At $t>t_{G}$ angle $\delta$ converges to either $0^\circ$ or $180^\circ$, as shown in Figure~\ref{fig:GravCapture}. These outcomes are equiprobable, the only difference between them being whether the rocket body orbits the Earth with thrusters down or up. Gravitational stabilized body rotates synchronously with the local vertical, therefore after $t>t_{G}$ angular venosity value tends to $\omega_o$ ($\tilde{\omega}\rightarrow 1$), as seen in Figure~\ref{fig:DeCap}.

The convergence of evolution to gravitational stabilization is a consequence of strong elongation of rocket body inertia ellipsoid, which leads to high gravity gradient torque. For debris objects with less elongated inertia ellipsoid, e.g less typical rocket bodies similar to Ariane 5 or defunct satellites, other final motion regimes exist (\cite{defsat2017}).

\subsection{Simulation results: evolution of angular momentum orientation}

To demonstrate all five scenarios of long-term evolution described in Section \ref{subsection:map}, numerical simulations for five sets of initial condition $N_1$--$N_5$ shown in Figure~\ref{fig:Map} were carried out. In all subsequent plots we shall use same colors as in Figures \ref{fig:CCycles}--\ref{fig:MSheet} to denote different regions corresponding to the system's state during the stage of exponential decay: orange~--~$R_M$, blue~--~$R_U$, green~--~$R_L$. Subsequent stages of slow chaos and gravitational stabilization are colored gray.

\begin{figure}[htb]
\center{\includegraphics[width=122mm]{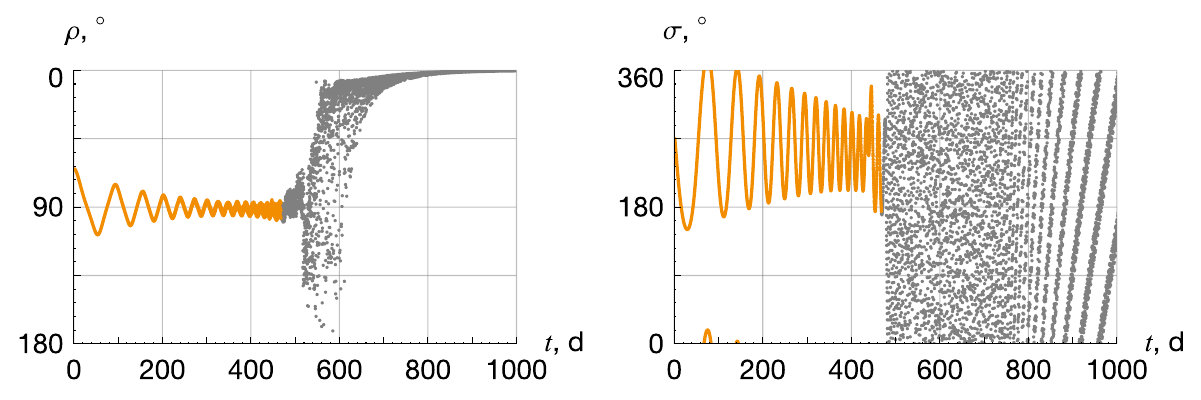}}
\caption{Evolution of angles $\rho$ and $\sigma$: an example of $R_M\rightarrow R_M$ scenario, based on set $N_1$ of initial conditions}
\label{fig:rhosigma(1)}
\end{figure}

Figure~\ref{fig:rhosigma(1)} shows the evolution of angles $\rho$ and $\sigma$ in case $N_1$, as an example of $R_M\rightarrow R_M$ scenario. The angular momentum vector here indeed oscillates about the direction to the south celestial pole as $\rho$ and $\sigma$ oscillate about values $90^\circ$ and $270^\circ$ respectively. The amplitudes of these oscillations decreases over time, as the system converges to Cassini state $P_M$ (Fig.~\ref{fig:CCycles}) in full agreement with Figure~\ref{fig:oscCC}.

\begin{figure}[htb]
\center{\includegraphics[width=122mm]{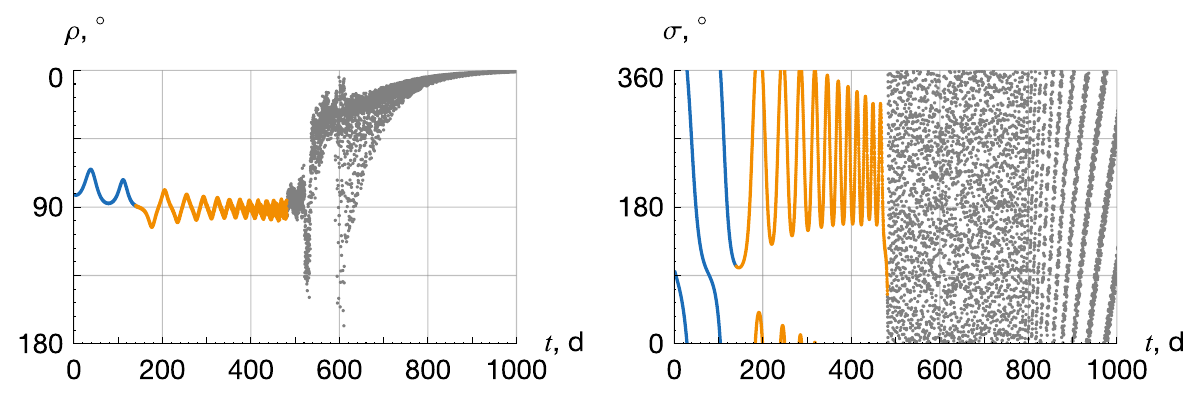}}
\caption{Evolution of angles $\rho$ and $\sigma$: an example of $R_U\rightarrow R_M$ scenario, based on set $N_2$ of initial conditions}
\label{fig:rhosigma(2)}
\end{figure}

In Figure~\ref{fig:rhosigma(2)}, which shows the simulation results for case $N_2$, the transition from region $R_U$ to $R_M$ at $t\approx140$ d is visible. As the angular momentum vector becomes captured in the middle region, the circulation of $\sigma$ over the whole interval $[0^\circ,360^\circ)$ is replaced by oscillation about $\sigma=270^\circ$.

\begin{figure}[htb]
\center{\includegraphics[width=122mm]{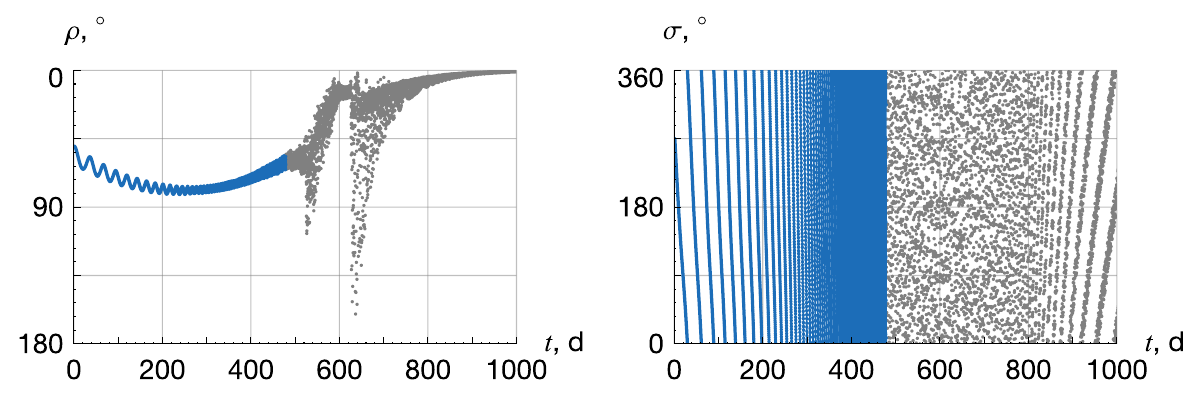}}
\caption{Evolution of angles $\rho$ and $\sigma$: an example of $R_U\rightarrow R_U$ scenario, based on set $N_3$ of initial conditions}
\label{fig:rhosigma(3)}
\end{figure}

Convex shape of plot $\rho$ in Figure~\ref{fig:rhosigma(3)} corresponds to the concavity of trajectories in $R_U$ in Figure~\ref{fig:oscCC} at $\tilde{\omega}\sim 30$. The angle $\rho$ increases while system comes closer to the separatrix, and starts to decrease, when it moves back to $P_U$. Thus, the axis of rotation in this scenario initially leans towards orbital plane, but deflects back to the orbital normal at $t\sim 300$ d. As explained in Section \ref{subsec:Rs}, this non-monotonous behavior is caused by the influence of the orbital motion on the eddy currents torque.

\begin{figure}[htb]
\center{\includegraphics[width=122mm]{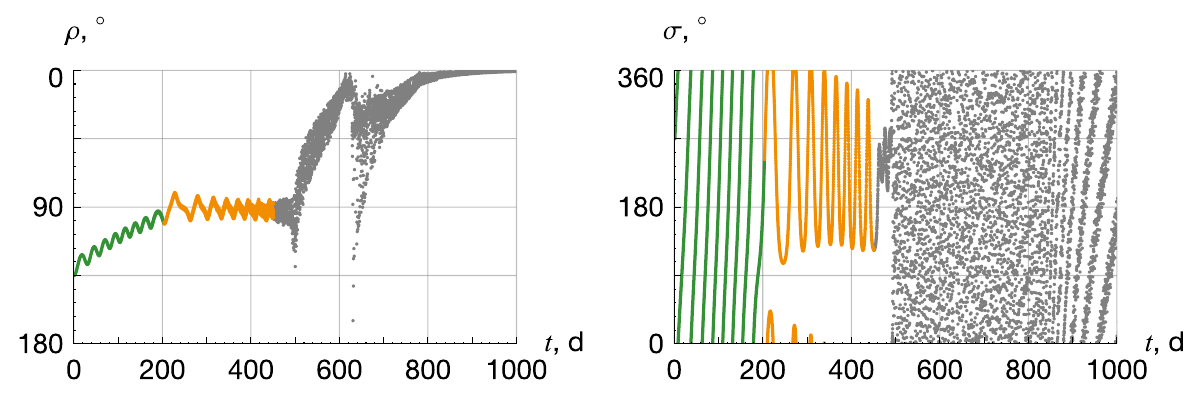}}
\caption{Evolution of angles $\rho$ and $\sigma$: an example of $R_L\rightarrow R_M$ scenario, based on set $N_4$ of initial conditions}
\label{fig:rhosigma(4)}
\end{figure}

\begin{figure}[htb]
\center{\includegraphics[width=122mm]{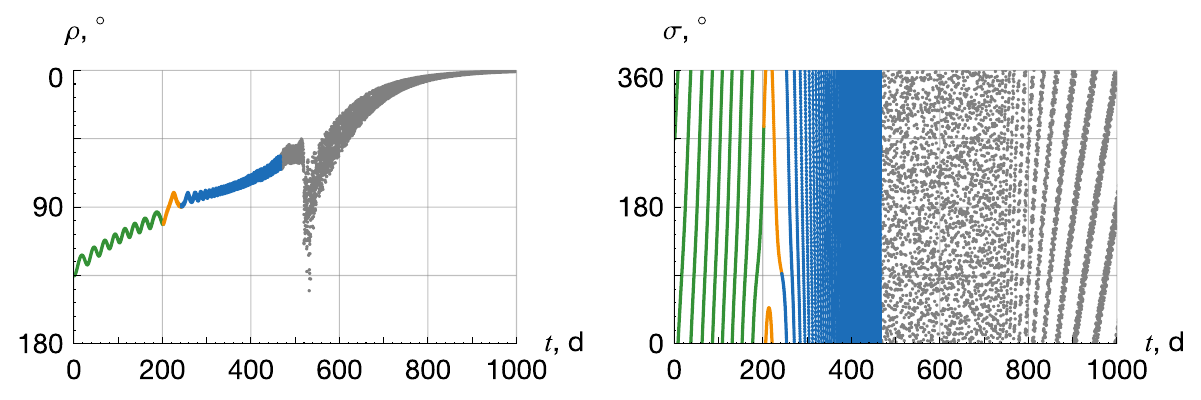}}
\caption{Evolution of angles $\rho$ and $\sigma$: an example of $R_L\rightarrow R_U$ scenario, based on set $N_5$ of initial conditions}
\label{fig:rhosigma(5)}
\end{figure}

Evolution for case $N_4$ is shown in Figure~\ref{fig:rhosigma(4)}. Here capture in the middle region of the trajectory that starts in $R_L$ is seen, which is very similar to case $N_2$. Alternatively, Figure~\ref{fig:rhosigma(5)} shows the evolution for case $N_5$ that starts very close to case $N_4$ (Fig.~\ref{fig:Map}), but instead of being captured into $R_M$, the systems jumps past it into $R_U$. After that, the angular momentum vector is carried away from the orbital plane towards $P_U$ by the orbital component of eddy currents torque, similar to the second half of evolution in case $N_3$. The transition $R_L\rightarrow R_U$ is also characterized by the change of angle $\sigma$ circulation direction (Fig.~\ref{fig:rhosigma(5)}).

Gray parts of plots in Figures~\ref{fig:rhosigma(1)}--\ref{fig:rhosigma(5)} correspond to evolution following the stage of exponential decay, and thus complement the analytical study carried out in Section \ref{sec:analytic}.
One can see, that in all cases after the end of the slow chaos stage, $\rho$ tends to $0^\circ$, as the gravitationally stabilized object rotates about the orbital normal.

\section{Conclusion}
Using analytical techniques and numerical simulation we have conducted a comprehensive study of the rotational motion of large objects in SSO. It is remarkable that despite of seeming insignificance, both precession of the orbit and influence of orbital motion on induced eddy currents proved to have a major impact on attitude dynamics.

The natural next step is to discover the predicted effects in the motion of the real objects. In particular, it would be desirable to check if the angular momentum vector (for objects similar to those we have modeled) in some cases indeed oscillates about the direction to the south celestial pole.

The other potentially observable phenomenon is the lack of fast rotating objects with retrograde spins (represented by region $R_L$ in our study). Most of them should fairly quickly switch to prograde spins, or become captured into angular momentum oscillations with the axis of rotation lying near the orbital plane.

One of the possible prospects of our work is the study of the rotational evolution of large space debris objects in the satellite class. In comparison with the rocket bodies the inertia ellipsoid of a typical satellite-like object is more similar to a sphere. Preliminary simulations show that among the final regimes for this class of objects there is not only the gravitational stabilization regime, but also rotation about the orbital plane normal with mean angular velocity equal to $9\omega_o/5$ ($\tilde{\omega}=1.8$), which is governed by eddy currents. In addition to that satellites might have a significant magnetic moment, which impacts the final stages of evolution and leads to even greater variety of final regimes (\cite{defsat2017}).

\begin{acknowledgements}
Research reported in this paper was supported by RFBR (grant 17-01-00902).
\end{acknowledgements}

\section*{Appendix}
This section provides the analytical expressions for integrals \eqref{eq:Ik} (\cite{B1954}). For $k\ge4$ the recurrence relation between $I_k$ is used:
\[
I_k = \frac{1}{{2(k - 1)}}\sum\limits_{j = 1}^4 {\left( {(2k - 2 - j){r_j}{I_{k - j}}} \right)}.
\]
Here $r_j$ are coefficients of $p^{4-j}$ in \eqref{eq:R4p}.

Calculation of the integrals $I_0\div I_3$ requires separate consideration of two specific cases.
\subsection*{Case 1: $R_4(p)$ has four real roots}
Here, as previously, $p_l$ is used to denote $R_4(p)$ roots ($p_1<p_2<p_3<p_4$). For integration over the interval $\left[p_1,p_2\right]$, which corresponds to Cassini cycle in $R_L$, the following relations hold:
\[{I_0} = g K(m),\quad
{I_1} = p_3 I_0 - ({p_3} - {p_2})g \Pi (n|m),\]
\[
{I_2} = ({p_3} + {p_4}){I_1} - {p_3}{p_4}{I_0} - g\frac{{({p_3} - {p_2})({p_4} - {p_2})}}{{2n(n - 1)}}\left[ {n E(m) + (m - n)K(m) + (2n - {n^2} - m)\Pi (n|m)} \right],
\]
\begin{multline*}
{I_3} = ({p_2} + {p_3} + {p_4}){I_2} - ({p_2}{p_3} + {p_2}{p_4} + {p_3}{p_4}){I_1} + {p_2}{p_3}{p_4}{I_0} + \\
 + \frac{{g{{({p_3} - {p_2})}^2}({p_4} - {p_2})}}{{8n{{(n - 1)}^2}(n - m)}}
 \left[ n\left(m(1 + 2n)- n(2 + n)\right)E(m) + \vphantom{\left( {(n - 4){n^3} + (1 - 4n){m^2} + 6{n^2}m} \right)} \right. \\
 \left.+ (m - n)\left(((4n - 1)m - n(n + 2)\right)K(m) + \left( {(n - 4){n^3} + (1 - 4n){m^2} + 6{n^2}m} \right)\Pi (n|m) \right].
\end{multline*}
Here $K(m)$, $E(m)$ and $\Pi(m)$ are complete elliptic integrals of the first, second, and third kind, and parameters $m$, $n$, $g$ are calculated as follows:
\[m = \frac{{({p_2} - {p_1})({p_4} - {p_3})}}{{({p_3} - {p_1})({p_4} - {p_2})}},\quad
n = \frac{{{p_2} - {p_1}}}{{{p_3} - {p_1}}},\quad
g = \frac{2}{{\sqrt {({p_3} - {p_1})({p_4} - {p_2})} }}.\]

Integration over the interval $\left[p_3,p_4\right]$, which corresponds to Cassini cycle in $R_U$, can be done by reflecting $R_4$ over axis $p$, so that the two greatest roots become the two smallest ones. In order to do that, the substitute $\left(p_1,p_2,p_3,p_4\right)\rightarrow\left(-p_4,-p_3,-p_2,-p_1\right)$ in the above formulae must be made and signs of integrals $I_0$ and $I_2$ changed.

\subsection*{Case 2: $R_4(p)$ has two real and two complex roots}

Here ${p_1},{p_2} \in \mathbb{R}^1$ ($p_1<p_2$), $p_{3,4}=\alpha\pm i \beta$ ($\alpha , \beta\in \mathbb{R}^1$) and the integrals are taken over the interval $\left[p_1,p_2\right]$ (region $R_M$ or $R_L$).
\[{I_0} = \frac{2}{{\sqrt {AB} }}K(m),\quad
{I_1} = \frac{{2(B{p_2} - A{p_1})}}{{\sqrt {AB} (B - A)}}K(m) + \frac{{(A + B)({p_1} - {p_2})}}{{\sqrt {AB} (B - A)}}\Pi (n|m),\]
\begin{multline*}
{I_2} = \frac{2}{{\sqrt {AB} (B - A)}}\left[ {\left( {A{p_1}^2 - B{p_2}^2 + 2\alpha \left( {B{p_2} - A{p_1}} \right) + \left( {A - B} \right)\left( {{\alpha ^2} + {\beta ^2}} \right)} \right)E(m)} \right. +\\
\left. { + \left( {B{p_2}^2 - A{p_1}^2} \right)K(m) + \frac{1}{4}\left( {A + B} \right)\left( {{p_1} - {p_2}} \right)\left( {{p_1} + {p_2} + 2\alpha } \right)\Pi (n|m)} \right],
\end{multline*}
\begin{multline*}
{I_3} = \frac{1}{{8\sqrt {AB} (B - A)}}\left[ 12\left( {A{p_1}^2 - B{p_2}^2 + 2\alpha \left( {B{p_2} - A{p_1}} \right) + (A - B)\left( {{\alpha ^2} + {\beta ^2}} \right)} \right)
({p_1} + {p_2} + 2\alpha )E(m)- \right.\\
- 4\left[ {A{p_1}^2(3{p_1} + {p_2} + 2\alpha ) - {p_1}\left( {B{p_2}^2} \right. + 2(A - B){p_2}\alpha  + (A + B)({\alpha ^2} + {\beta ^2})} \right)+\\
\left. { + {p_2}\left( {(A + B)({\alpha ^2} + {\beta ^2}) - B{p_2}(2\alpha  + 3{p_2})} \right)} \right] K(m) + \\
\left. { + (A + B)({p_1} - {p_2})\left( {3{p_1}^2 + 2{p_1}{p_2} + 3{p_2}^2 + 4({p_1} + {p_2})\alpha  + 8{\alpha ^2} - 4{\beta ^2}} \right)\Pi (n|m)} \right].
\end{multline*}
Parameters
\[m = \frac{{{{({p_2} - {p_1})}^2} - {{(A - B)}^2}}}{{4AB}},\quad
n = \frac{{{{(A - B)}^2}}}{{4AB}},\quad
A = \sqrt {{{({p_2} - \alpha )}^2} + {\beta ^2}} ,\quad B = \sqrt {{{({p_1} - \alpha )}^2} + {\beta ^2}} .\]

\end{document}